# Profiling the Cybercriminal: A Systematic Review of Research

Maria Bada
*University of Cambridge, UK*
Maria.Bada@cl.cam.ac.uk

Jason R.C. Nurse
*University of Kent, UK*
J.R.C.Nurse@kent.ac.uk

*Abstract*—As cybercrime becomes one of the most significant threats facing society today, it is of utmost importance to better understand the perpetrators behind such attacks. In this article, we seek to advance research and practitioner understanding of the cybercriminal (cyber-offender) profiling domain by conducting a rigorous systematic review. This work investigates the aforementioned domain to answer the question: what is the state-of-the-art in the academic field of understanding, characterising and profiling cybercriminals. Through the application of the PRISMA systematic literature review technique, we identify 39 works from the last 14 years (2006-2020). Our findings demonstrate that overall, there is lack of a common definition of profiling for cyber-offenders. The review found that one of the primary types of cybercriminals that studies have focused on is hackers and the majority of papers used the deductive approach as a preferred one. This article produces an up-to-date characterisation of the field and also defines open issues deserving of further attention such as the role of security professionals and law enforcement in supporting such research, as well as factors including personality traits which must be further researched whilst exploring online criminal behaviour. By understanding online offenders and their pathways towards malevolent behaviours, we can better identify steps that need to be taken to prevent such criminal activities.

## I. INTRODUCTION

Cybercrime and the persistent wave of cyber-attacks being witnessed in society today pose significant challenges to security in the 21st century. To properly examine the topic of cybercrime, it is important to first define it. For our purposes, we use the definition provided by the European Commission (EC) given its broad nature and thus, ability to allow us to adopt a more inclusive approach to our work. As such, cybercrimes hereafter are regarded as "*criminal acts committed using electronic communications networks and information systems or against such networks and systems*" [17]. This definition encompasses a large range of crimes (e.g., hacking, social engineering, fraud, denial of service attacks and online harassment) and has been used as the foundation for several other academic research works [1], [6], [42].

Across government, industry and academia, there have been numerous efforts to address the issue of cybercrime. Governments have largely responded with regulation and bolstered law enforcement capabilities (e.g., The Joint Cybercrime Action Taskforce (J-CAT) of Europol), industry has prioritised security policies, procedures and controls, and academia has pursued a combination of approaches (across domains such as criminology, computing and psychology). To complement these proposals, cybersecurity technologists have offered an array of platforms and tools.

While the efforts above are crucial parts required to address the problem, another component that has received considerably less dedicated attention is that of developing a comprehensive understanding of cybercriminals (hereafter also referred to as cyber-offenders) themselves, i.e., the notion of 'profiling cybercriminals'. Criminal or offender profiling, in general, is a tool that has been used by forensic experts for decades and can help to identify the offender's behavioural tendencies, personality traits, demographic variables, and geographical variables based on the information and characteristics of the crime [37]. Applying such techniques to cyber-offenders could offer similar benefits to the online space, and act as another tool for law enforcement to draw upon.

This article furthers research and practitioners' understanding of the cybercriminal-profiling domain by conducting a comprehensive systematic review. Systematic reviews expand on the concept of traditional literature reviews by mandating a series of systematic, structured and explicit methods to gather and analyse relevant research [41]. In particular, this work synthesises and critically reflects on the aforementioned domain to answer the question: what is the state-of-the-art in the academic field of understanding, characterising or profiling cybercriminals? Through this analysis we expect to extract crucial insight into this domain, its nature and development over several years.

Specifically, we define the most common venues (e.g., journals, conferences, archives) where articles in this field are disseminated, the domains in which their authors originate (e.g., through academic affiliation, or professionals in law enforcement or industry) and the general types of studies conducted (e.g., reviews, profiles, methodologies). This work further investigates the nature of the articles and the contributions they seek to make to the body of knowledge of understanding cybercriminals, their actions and behaviour. These contributions, as an example, may pertain to the definitions of profiling/understanding, variety of datasets used/analysed, proposed profiles themselves (deduced from research or practice), and approaches (or methodologies) to better understand cybercriminals and characterise them (e.g., motivations, etc.). In addition to forming a comprehensive understanding of the articles, we are also interested in whether there is any demonstrated or intended engagement with law enforcement agencies (LEAs); these are the ideal users of such research. Through the investigation of the various aspects mentioned above, we are able to provide in-depth insights into current research and also identify gaps in knowledge in the topic space, and therefore highlight where further research within this domain is required.

## II. METHOD

### A. Protocol and registration

The core value of the systematic review process is in the structure that it provides to the search, gathering and analysis of pertinent literature. To guide our study's application of this process further, we have chosen the PRISMA (Preferred Reporting Items for Systematic reviews and Meta-Analyses) technique and protocol [41]. PRISMA outlines a minimum set of items/tasks required for conducting and reporting systematic reviews. It is evidenced-based, well-regarded and straightforward to apply. Our adoption of this technique involved defining each of the items necessary and creating an appropriate plan of analysis. For instance, this entailed identifying eligibility criteria for the literature search, relevant information sources, and the data collection process. The main sections (and subsections) that follow each map to items outlined by PRISMA.

### B. Eligibility criteria

To guide the selection of articles of the review, there were two inclusion criteria. The first (IC1) required articles to be published/reported in English, and the second (IC2) sought to elicit articles that pertain to the understanding, characterising or profiling of cybercriminals or cyber-offenders. While IC1 was imposed due to the language spoken by this article's authors, we believe this is a reasonable criteria given English is widely regarded as the primary language of the community. The rationale for IC2 can be found in this study's research question and the direction it has outlined for our review. It should be noted that this study does not consider articles that deal with the understanding or profiling of traditional criminals conducting crimes offline. Crimes, and therefore articles, that would be of interest are those which fit the description of cybercrime presented in Section I.

### C. Information sources

The research topic under investigation stretches across numerous domains and therefore a wide range of information sources were searched for appropriate articles, including journals, conference proceedings, and book chapters. Publication databases explored include well-known and reputable sources, namely, Scopus, Web of Science, ProQuest, ACM Digital Library, IEEE Digital Library, ScienceDirect, and Springer. In addition to the articles discovered in the various (academic and otherwise) searches above, the reference lists of the articles themselves were (iteratively) analysed to identify relevant, new contributions. All searches were conducted in June 2020.

### D. Search and article selection

The search and article selection process followed steps informed by the factors defined earlier. The first step was searching for articles in the databases listed that were relevant to the aims of this research. Search queries were constructed using key words commonly associated with the general research topic. To capture the criminal aspect, we used: "cybercriminal" and "cyber-offender" (in addition to appropriate variations such as "cyber-criminal" and "cyber offenders"). To accommodate for the profiling component, we used a wide range of keywords including: profiling, behaviour, methodology, psychological and personality. Tests with keywords were conducted and refined to ensure that they captured a wide set of articles—if articles were later found to be irrelevant, this was preferred to potentially excluding relevant articles.

The final keyword search query was as follows: *(("cyber criminal" OR "cyber criminals" OR cybercriminal OR cybercriminals OR "cyber offender" OR "cyber offenders") AND (understanding OR profiling OR profile OR behaviour OR behavior OR characterising OR methodology OR psychology OR psychological OR personality))*.

The discovered articles were then examined according to the eligibility criteria, with special attention first placed on their titles, abstracts and key word lists, and then as necessary (e.g., if the article appears to fulfil the first two criteria) on the full text. As articles were reviewed, attention was also placed on their reference lists and bibliographies, in the event that additional relevant work was present (and therefore should be included). To maintain a rigorous approach to the research, the broad search and selection process was undertaken independently by this paper's two authors; differences on the selection process of additional relevant work were discussed and agreed upon. For example, the authors excluded some papers which were out of scope for this work, since they didn't focus on profiling, but in other topics in relation to cybercrime [40]. Before continuing, we note that the decision to use the word 'cyber' as opposed to 'online' or 'internet' was intentional. This was done to focus our study on research that considered their work on cyber-criminal/offender profiling. Moreover, this decision was in line with our more broad emphasis on cybercrime. We discuss the implications of this approach later in the article.

### E. Data collection process and Data items

From the selected articles, relevant data was collected using a data extraction form. This form outlined key features of interest which aligned with the systematic review's aim. Each article was assessed separately by the two authors and differences in the data extracted were discussed for an agreement to be made.

To structure our data extraction task, we specified variables appropriate to this aim. These functioned at two levels: descriptive—to capture high-level metadata about each article and thus assist in supplying a overview of the field of literature; and in-depth—to focus on more detailed analyses and thereby develop a detailed understanding of the manuscripts. These variables are presented in Table I.

| | |
|---|---|
| Descriptive | Publication or dissemination year, subject area, venue type (e.g,. journal, conference, report) and venue (e.g., name of journal) |
| | Affiliations of authors, including location, subject area, profession (e.g., academic, law enforcement, industry) |
| | Contribution type (e.g., reviews, proposing/exploring profiles, proposing methodologies) |
| | Use of data, data gathering approach (e.g., interviews, questionnaires, or online data from websites, forums, social media, etc.) and data analysis approach (e.g., content analysis, thematic analysis, machine learning, text analysis, etc.) |
| In-depth | Definition of cybercriminal understanding or profiling |
| | Type(s) of cybercriminal(s) in focus and their nature |
| | Approach to understanding or profiling |
| | Contributions of the article |

TABLE I
DATA ITEMS USED TO EXAMINE SELECTED STUDIES

Having specified the structure of our systematic review, including articles of interest and key data to be extracted from selected articles, next we present and discuss the results.

III. RESULTS

A. Selection of articles

In total, 39 articles were selected for our review. From our database search, we initially found 3189 articles across the seven sources. After deduplication, 1023 were removed leaving 2166 remaining. Application of IC1 resulted in us discarding 3 manuscripts, while application of IC2 (which also included reviewing full text as necessary) removed 2127 articles. Reference lists of the remaining articles were examined and 3 articles meeting IC1 and IC2 were included. This resulted in 39 studies for this review.

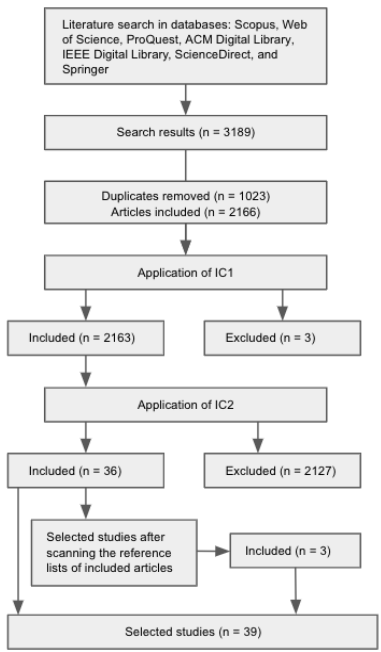

Fig. 1. PRISMA flow diagram

The process adopted is also depicted in Figure 1.

B. Characteristics of articles

The articles selected from our review provided key insights into the cybercriminal profiling domain, including its nature, and how research efforts have progressed over time. In this section we focus on descriptive characteristics of these articles.

Across all years, there were 24 articles in journals, seven in conferences and nine in book chapters. From our analysis we found only six sources appear more than once: Crime, Law and Social Change Journal (3), Digital Investigation Journal (2), Trends in Organized Crime (2), Information Systems Frontiers (2), and the European Conference on Cyber Warfare and Security (2).

Reflecting on the location of authors that have contributed to the field of profiling, 16 countries are represented (and we state the number of articles in brackets): USA (12), UK (10), The Netherlands (6), Australia (4), South Africa (2), Switzerland (2), Estonia (1), Brazil (1), Bosnia and Herzegovina (1), Spain (1), Canada (1), Ghana (1), China (1), Philippines (1), Nigeria (1), and South Korea (1).

On the topic of cybercriminal profiling specifically, a number of studies contributed either in methodologies for profiling [9], [29], [46], [55], in the application of profiling in areas [5], [8], [18], [25], or in open issues in profiling [4].

The topic of cybercriminal groups featured in numerous studies, as authors sought to understand and characterise the behaviour of these groups. Contributions engaged in general analysis of groups including those involved in organised crime [7], [12], [13], [32], [33], [35] while further investigating questions such as how they build trust [59] and what typologies of cybercriminal networks may exist [34].

The type of data used to investigate cybercriminals is also crucial to understanding the nature of this research on this topic. We found that 22 (or, just over half of the) articles rely on new data, be it from surveys, interviews (with criminals and law enforcement), court records, case studies, underground markets or honeypots [3], [7], [8], [10], [11], [24], [29], [31]–[36], [43]–[45], [48], [49], [52], [57]–[59]. The other studies primarily use existing literature to form their arguments and often rely on research from other fields, or established approaches mapped to the cybercrime domain [2], [4], [5], [9], [12], [13], [15], [18], [21], [25], [27], [46], [50], [53]–[56]. This balance is interesting to note as it highlights the extent of articles on profiling that are not working with primary data.

C. Definitions of cybercriminal profiling

Most of the papers we reviewed included a definition on criminal profiling or offender profiling rather than cybercriminal profiling. For example:

> *Criminal profiling is an investigative approach based on the assumption that the crime scene provides details about the offense and the offender [8].*

One paper provided a specific definition of cybercriminal profiling [10]. That paper also focused more on the motivations behind engaging in cybercrime. As described: "*....Determining and understanding why hackers do what they do*". Overall, the definitions provided focused on different criminal profiling approaches such as inductive and deductive profiling [3], [4], [9], [11], [46], [48], [55], geographic profiling [8], and psychological profiling [46].

*D. Types of cybercriminal(s) researched*

One of the aspects of interest in this work is the type of cybercriminals studied in cybercriminal profiling research.

The papers focusing on hackers study white hat, black hat hackers and hacktivists [3], [4], [7], [12], [13], [18], [21], [27], [49], [54]–[56]. The basis on which different members of the hacker community differentiate themselves is the ethics of their activities, such as black hat or white hat hackers which acknowledges the use of malicious or ethical applications of hacking [20]. Hacktivists can be motivated by political views, cultural/religious beliefs or terrorist ideology [19]. Also, the focus seems to be on hackers using carding forums. An example of a carding forum is Tor which enhances the anonymity of user activity on the dark web and enables criminals to quickly distribute compromised information and to profit from using it in a fraudulent way [30]. As it pertains to fraudsters, studies focused on counterfeits and reproductions, in fraud and/or unauthorised access as well as cybercrimes affecting the banking sector.

Overall, the majority of studies seem not to focus on specific types of cybercriminals, rather employ a general lens while looking at the topic. In doing so, these studies focus on factors such as educational-attainment, modus-operandi, and networks-collaborators [31]. As stated in one paper: *Cybercriminals are a diverse group, yet you can spot some patterns and motives across the group* [52]. Finally, some papers attempted to construct typologies for organised cybercriminal networks [12], [24], [34]. For example, different roles were identified within cybercriminal phishing networks: core members, professional enablers, recruited enablers, and money mules [34].

*E. Approaches to profiling or understanding cybercriminals*

We now move to describe the different approaches adopted to profile or understand cybercriminals, based on the studies in our review. By 'approach' we refer to the different frameworks or models utilised to studying profiles and the methodologies followed in the analysis.

Through this assessment, our work identified several different methods and approaches used to construct profiles or understand online offenders. Some papers were descriptive in their approach, while others were based on case study analysis [7], [12], [13], [21], [54]. Others conducted interviews in order to construct profiles and collect information about the personality and practices of online offenders [31], [33], [35]. In terms of the method used to analyse information, approaches such as inductive and deductive profiling were found [4], [5], [9], [48], [55]. As described:

> *Inductive profiling involves the study of a group of subjects who share a common characteristic or activity to discern trends or patterns in their motives, characteristics or behavior. Deductive profiling refers to the assessment of a subject's personal characteristics from his or her crimes, activities, statements or other reports and is associated with case investigations [48].*

For some papers, the deductive approach was preferred [5], [9], [48]. The deductive approach explores information such as demographics, personal characteristics, risk related behaviour, impulsiveness, history and personality traits. It has been used, for instance, in order to study the profiles of insider threats [48]. It appears to be a preferred method for many since it is more applicable to case-based studies, in which the crime, attack or threat is analysed and specific characteristics of the perpetrator are developed [7], [12], [13], [21], [54].

Some studies have stepped further and proposed a profiling methodology incorporating both the inductive as well as the deductive profiling approaches [5], [55]. For example, the proposed methodology for cyber criminal profiling by [55] employs six Profile Identification Metrics to determine the offender's modus operandi, psychology, and behavior characteristics. These are: attack signature, attack method, motivation level, capability factor, attack severity, demographics. Similarly, a generic, deductive-oriented profiling framework with a dominant behavioural evidence analysis influence is being suggested in another study [5]. This framework is integrated to a digital forensics process methodology, and in addition to the technical and behavioural aspects, a legal aspect is introduced to guide the process of profiling where issues of transnationalism are encountered.

Another approach identified is the behavioural analysis used to analyse online auction fraud offenders, based on personal characteristics and motivation data [11], [46]. This approach is often described as:

> *Psychological profiling and investigative psychology have played and will continue to play a crucial role in understanding why certain crimes are committed, developing profiles of likely suspects, and linking crimes to specific individuals or groups [46].*

Geographic profiling is another approach used for cybercrime investigations [8].

The researchers developed GeoCrime, a geographic profiling software designed to assist in the mapping, spatial and statistical analysis of cybercrime patterns. Geographic profiling is important, especially in situations where little is known about the offender, such as in cybercrime, where offenders use the Internet to hide their identities and activities.

Also, other work has used psycho-linguistic analysis [9] and crime script analysis [56] methods. A crime script analysis involves breaking down the actions of the criminal into four main stages: preparation, pre-activity, activity, and post activity. As this study claims, an appreciation of the offender's mind/perspective seems to be of the greatest importance in the creation of crime scripts, to better understand the behaviours, motives, feelings, decisions within the process of a crime [56].

The well-known criminal profiling framework of the FBI [16] was also used to collect and analyse information [3], [44]. The FBI criminal profiling model follows a deductive and inductive approach. It involves the analysis of evidence at one particular crime scene to build possible offender characteristics, based on the profiler's experiences and expertise. The inductive profiling involves exploring detailed data on past crimes committed of a certain type to establish correlations, patterns, and similarities largely shared between the characteristics, motivations and methods of offenders committing a specific type of crime [3].

Finally, only one paper proposed a model for profiling cybercriminals incorporating elements from different approaches of profiling (deductive, inductive, behavioural analysis) [5].

*F. Primary research contributions in the field*

Upon reviewing the selected studies, the majority provide different types of research contributions on existing knowledge in the field. For a number of articles the description of different types of offenders and the different ways they exploit ICTs (e.g., traditional organised criminal groups, organised cybercrime groups, and ideologically and politically motivated cybercrime groups) is the main focus [12], [25], [31], [49].

Along the same area of research other studies explored the organizational structures of groups involved in cybercrime [32]–[35]. For instance, Leukfeldt et al. [34] discuss the differences distinguished between networks carrying out low-tech attacks and high-tech attacks as well as the different layers of these networks such as the core members (the higher lever), professional enablers (a layer below the core members) and money mules (the bottom layer). Other studies approached the topic from a different angle, analysing carding forum data to explore the ways cybercriminals communicate and build trust [59] and the way they network and cooperate with each other to commit offences [7], [24].

Key findings were also found in the area of the description of cybercrime and cybercriminals, cultural constructions of cybercrime and cybercriminals or the culture of fear [9], [54]. Another important area of significant findings within this review was on the differences between online and offline offenders [58] as well as the exploration of online criminal behaviour by applying social and criminological theories related to cybercrime [53].

However, another study [53] highlighted the differences of the characteristics of online crimes and conventional crimes such as proximity of the criminal to the victim and crime location as well as the scale and size of the crime that can be committed which is limited by physical constraints.

The motivating factors behind online criminal behaviour was a central component of a number of studies [3], [10], [45], [52]. These studies have identified that factors such as political ideology, economic or social factors or the feeling of personal accomplishment which can lead to online criminal behaviour. Only limited work provided further contributions in understanding cyber-offenders by exploring demographic and contextual attributes [43], [57]. In Payne et al. [43] for example, they found that online offenders may be younger than all other offender types, and they are usually male or are likely to be white.

Overall, the main aim of most studies has been to identify the trends of cybercrime and the strategies employed by cyber-offenders to commit cybercrime in order to identify appropriate steps that need to be taken to prevent such criminal activities.

IV. DISCUSSION

The main purpose of this study is to explore research relevant to profiling of cybercriminals. In particular, this work synthesises and critically reflects on work around understanding, characterising or profiling cybercriminals. In this section, the main points that have emerged from reviewing the articles surveyed will be discussed below.

*A. Study characteristics*

It is not a surprise that a large proportion of studies focused on individual cases of cybercriminals for their analysis [24], [45], [50], [52], [53]. For example, the analysis of cases brought before the courts might not be representative of the larger population of cyber-offenders who are not prosecuted. However, gaining data by interviewing large numbers of such offenders is extremely challenging.

In addition, studies have assessed different types of cybercriminals including hackers [3], [10], [15], [21], [27], [49], [54], fraudsters [2], [11], [31], [36] and insider threats [48]. Another interesting finding is that many studies [2], [4], [5], [9], [12], [13], [15], [18], [21], [25], [27], [46], [53]–[56] do not work with primary data but rather rely on research from other fields adapting it to the cybercrime domain. This is quite interesting because it illustrates the difficulty in conducting research in the field of cybercrime due to the lack of relevant data. It also creates a barrier to entry for new researchers who are unable to conduct research in the field due to the inability to secure data access.

The lack of data might lead to incorrect assumptions especially regarding the psycho-social characteristics of offenders. In order to bridge this gap, academics need to align their efforts with law enforcement agencies and work towards collaborations which allow secure, ethical and legal access to relevant data. In terms of the topics of focus that these studies explored, we noted that often a comparison between online and offline offenders is conducted to identify differences or similarities [?], [43], [58]. In addition, topics such as criminal motivations, behaviours, relationships between crime and developmental disorders, mapping criminal journeys are often in focus.

*B. Definitions of cybercriminal profiling*

Regarding defining cybercriminal profiling, only ten of the 39 studies reviewed provided a definition of what profiling is grounded in the approach that they are using.

Overall, there is a richness of definitions on profiling for cyber-offenders due to the differences in motivations, skill levels or cultural contexts. This lack of homogeneity in the population of criminals might also justify the complexity of exploring this field. These reasons have resulted in a literature that is still maturing, with more work needed to understand the links between skills, personality traits and motivation behind online criminal behaviour.

In addition, a number of personality or crime theories have been utilised to understand and analyse criminal behaviour however that is not the case for online offender behaviour. For example, a significant amount of work in psychology has been conducted in relation to personality traits and crime. Eysenck's Theory of Crime [22] proposes a three-dimension model (PEN) of personality: Psychoticism (anti-social, aggressive and uncaring), Extraversion (sensation-seeking) and Neuroticism (instability). These three personality dimensions form a unique set of characteristics that renders an individual susceptible to criminal behaviour. Also, given the importance of the Dark Triad and traditional crime, the Psychoticism, Narcissism and Machiavellianism traits have also been linked to criminal behaviour [47].

*C. Types of cybercriminal(s) researched*

The review found that one of the primary types of cybercriminals that studies have focused on is hackers. One example of which is a study exploring black hat hackers and offering valuable new insights on the psychological reasoning in the hacker's decision-making process, during their crime life cycle [49].

Furthermore, the majority of studies employ a general lens while looking at the topic. In doing so, studies focus on factors such as educational-attainment, modus-operandi, and networks-collaborators [31], [32], [34], [35]. This general approach once again stresses the need for more targeted research in order to identify differences among different types of offenders, informing law enforcement prevention activities. Another focus in the articles reviewed is also how cybercriminals communicate and build trust [59]. The findings indicate that carding forums facilitate organised cybercrime because they offer a hybrid form of organisational structure that is able to address sources of uncertainty and minimise transaction costs to an extent that allows a competitive underground market to emerge. Therefore, further research is necessary on different characteristics of carding forum members such as skills, motivation but also links to offline crime.

*D. Approaches to profiling or understanding cybercriminals*

In terms of the method, a broad range of methods were identified such as: a) interviews; b) case studies; c) psycholinguistic analysis of digital communications; d) crime script analysis; e) behavioural analysis; f) deductive or inductive methodology to analyse the information; g) FBI's criminal profiling framework; h) mathematical framework, constructing typologies for organised cybercriminal networks; and i) geographic-mathematic framework analysis.

However, the majority of papers used the deductive approach as a preferred one. This is mainly due to the fact that it is more applicable to case-based studies, in which the crime, attack or threat is analysed and specific characteristics of the perpetrator are explored. That is due to the lack of large available data-sets to be analysed systematically or over time. For example, Action Fraud in the UK is collecting information on victims of cybercrime however, the focus in not on offenders.

As mentioned above, one of the most well-known approaches in profiling is the FBI's criminal profiling framework. This approach has been applied to profile hackers, by analysing information such as the date of incident, demographic characteristics of the hacker, motives, presence of an accomplice, attack tools, and whether the attack happened locally only or internationally [3]. Although, this approach is being applied to provide insights on online behaviour, still the characteristics of the hacker are mainly demographics, lacking the social and psychological insights that various studies have been exploring for decades in offline criminal behaviour.

It is also identified that researchers are selecting some parts of specific frameworks which they find appropriate to analyse the type of data they have collected or analysing. For example, a study [32] has selected some aspects of an analytical framework [28] used to systematically analyse cases of organised crime. The selected aspects of the framework were ties between members of networks, processes of origin and growth, the composition of the network, its structure, the dependencies, the background of members, social ties and the (digital) offender convergence settings used by the criminals.

*E. Primary research contributions in the field*

*1) Data used:* Studies followed different methods in data collection in examining the profiles of cyber-offenders. Some studies have attempted to construct typologies for organised cybercriminal networks, other studies explored the organizational structures of groups involved in cybercrime, while others analysed carding forum data to explore the ways cybercriminals communicate and build trust or the way they network and cooperate with each other to commit offences.

*2) Working with law enforcement:* It has also been identified that for some studies the engagement with law enforcement was necessary. Some researchers have interviewed law enforcement officers and detectives [24], [31], [33]–[35] as a way of capturing the experience and knowledge of computer crime or fraud specialists. Other studies have focused on analysing police investigation files [35]. The role of law enforcement is really important is conducting research in this field. For example, the UK's National Crime Agency supported research exploring autism and cyber-dependent crime [44].

*3) Theoretical approaches in profiling:* An important area of significant contributions in the field is the study of the differences between online and offline offenders [58] as well as the exploration of online criminal behaviour by applying social and criminological theories related to cybercrime [53].

We found however that there is very limited work looking at personality aspects in relation to cybercrime. The main contribution in understanding online offenders is mainly done by exploring demographic and contextual attributes as well as motivations [43]. In Payne et al. [43] for example, they found that online offenders may be younger than all other offender types, and they are usually male or are likely to be white. However, studies on offline crime have pointed to personality as being a significant predictor of criminality, estimated at twice the size of the more commonly touted factor of social class background effects [51]. For example, a study [14] applied the Five Factor Model identifying agreeableness and conscientiousness as additional personality factors which should be considered when building a personality profile of offline criminals. In addition, low self-control has been linked to online criminal behaviour as it characterises individuals being impulsive, insensitive, risk-taking and fail to consider the long-term consequences of their actions [23]. Also, studies have linked personality to insider threat cases [39].

*4) New method development:* There are only a small number of studies suggesting new methods or developing a framework to profile online offenders incorporating both the inductive as well as the deductive profiling approaches [5], [55]. However, only one study from the ones selected for this review proposed a model for profiling cybercriminals incorporating elements from different approaches of profiling (deductive, inductive, behavioural analysis) [5]. In the past studies have suggested theoretical models such as the theoretical model profile of a hacker [38] aiming to indicate relations between the offender's characteristics, the environment and her/his success and the modus operandi during the attack.

*5) Gaps and policy implications:* The findings from the studies reviewed have identified the current gaps around understanding the psychology of online criminal behaviour and the prevention of cybercrime and stressed potential policy implications [21]. It has been identified that there is a need for new strategies of response and further research on analysing organised criminal activities in cyberspace [12]. In addition, the fact that digital evidence will be used more in judicial proceedings in the future, stresses the need for law enforcement to have in-depth knowledge of computer forensic principles, guidelines, procedures, tools, and techniques [12].

Another identified aspect for further research is the study of the characteristics of both the offender and offense to predict the societal reaction of the jurisdiction system on computer assisted frauds. Examining criminal capacity and culpability and identifying how the characteristics of the offender and offense itself might affect the application of law has been identified as something currently lacking [36].

## V. Conclusions and Future Work

Our review considered the salient characteristics of current studies, information on cybercriminals, their actions and behaviour, and crucially, the methods used to profile and understand such offenders. Based on our review, it is clear that there is steady growth in research into cyber-offender profiling. There are, however, some areas where research is lacking. We discuss these below.

*Common definitions*: Future research should be looking more closely on accepting a common definition on cybercriminal profiling. One example which may form the basis for future discussions is *"cybercriminal profiling is an educated attempt to provide specific information as to the type of individual who committed a certain crime. A profile based on characteristics patterns or factors of uniqueness that distinguishes certain individuals from the general population"* [55].

*Common Approach*: Future research should give sufficient attention on a more common approach of cybercriminal profiling so that it can be tested and applied in a systematic way. This way more standardised methods can be developed specifically for studying online criminal behaviour.

*Personality traits and online criminal behaviour*: Factors including personality traits must be further researched whilst exploring online criminal behaviour and how they influence the pathway a person can choose towards delinquency or not.

*Lack of data*: The difficulty in conducting research in this field is the lack of large data sets that can be analysed as well as the lack of access in such data. In addition, currently there is lack of relevant data collected by law enforcement agencies. Future studies would focus more on establishing collaborations with law enforcement agencies, informing their practises and data collection approaches.

*Wider focus*: Future research looking at profiling of cybercriminals should employ a wider focus of types of offenders. Currently, there seems to be a lot of research focused on hackers or cyber crimes in general, but not on other specific types such as on hacktivists [26]. Future studies could also address the limitation of this study, by including non-academic sources in future analysis or examining potential geographic discrepancies in profiling definitions and methodology used.